\documentclass{article}
\usepackage{amsmath}
\usepackage{amssymb}
\usepackage{graphicx}
\def\beq{\begin{equation}}
\def\eeq{\end{equation}}

\newcommand{\ed}{\end{document}}

\begin{document}

\title{Spherical Solutions due to the Exterior Geometry of a Charged Weyl Black Hole }

\author{Farrin Payandeh$^1$\thanks{e-mail: f$\_$payandeh@pnu.ac.ir} ,  Mohsen Fathi$^{2}$\thanks{Corresponding author: e-mail:
mohsen.fathi@gmail.com}}

\maketitle   \centerline{\it $^{1}$Department of Physics, Payame
Noor University, PO BOX 19395-3697, Tehran, Iran}
\centerline{\it$^2$Department of Physics, Islamic Azad University,
Central Tehran Branch, Tehran, Iran}

\begin{abstract}
Firstly we derive peculiar spherical Weyl solutions, using a
general spherically symmetric metric due to a massive charged
object with definite mass and radius. Afterwards, we present new
analytical solutions for relevant cosmological terms, which appear
in the metrics. Connecting the metrics to a new geometric
definition of a charged Black Hole, we numerically investigate the
effective potentials of the total dynamical system, considering
massive and massless test particles, moving on such Black Holes.
\end{abstract}

\section{Introduction}

Among generalized theories of gravity, Weyl gravity is remarkable,
since it leads to considerable descriptions of cosmological
parameters relevant to Dark Energy problem. It is known that Weyl
theory, contributes in the $R^2$ theories of gravity. This means
that this theory is governed by field equations, combined of
second order differentiations of the Ricci scalar
\cite{Mannheim1}:
$$W_{\alpha\beta}=\nabla^\rho\nabla_\alpha
R_{\beta\rho}+\nabla^\rho\nabla_\beta R_{\alpha\rho}-\Box
R_{\alpha\beta}-g_{\alpha\beta}\nabla_\rho\nabla_\lambda
R^{\rho\lambda}$$
\begin{equation}
-2R_{\rho\beta}
R^{\rho}_\alpha+\frac{1}{2}g_{\alpha\beta}R_{\rho\lambda}R^{\rho\lambda}-\frac{1}{3}\Big(2\nabla_\alpha\nabla_\beta
R-2g_{\alpha\beta}\Box
R-2RR_{\alpha\beta}+\frac{1}{2}g_{\alpha\beta}R^2\Big)=\frac{1}{4\pi}T_{\alpha\beta}.
\label{Weyl main}
\end{equation}
In the above equation, $W_{\alpha\beta}$ notates the components of
the Weyl equations, and $T_{\alpha\beta}$ is the energy momentum
tensor, corresponding to the source. The vacuum equations,
generally have been solved and a spherically symmetric metric has
been derived \cite{mannheim2}. In this paper, we are not
concerning about the general solution. Instead, we consider a
peculiar one, that corresponds to a massive charged spherically
symmetric source. Such source has been considered for the
Reissner-Nordstr\"{o}m solutions (RN) of Weyl gravity
\cite{mannheim3}, but here we use the background field method and
linear approximation to derive other RN-like solutions, also
describing a charged Black Hole. The method is like the one, which
has done in \cite{Tanhayi}, analytically investigating the
constants regarding the Dark Energy problem. However here,
constant values of charge and mass will be considered, to
rearrange the metric to a new form to describe a Black Hole.
Finally, the effective potentials will be numerically plotted to
illustrate the geometric behavior of the dynamical system,
affected by such Black Hole.

\section{A spherical solution of Weyl field equations due to a charged massive spherical source}

Let us consider the following general metric:
\begin{equation}
ds^2 = -\Big(1-\frac{2GM}{r}-\frac{1}{3}f(r)\Big)dt^2 +
\Big(1-\frac{2GM}{r}-\frac{1}{3}f(r)\Big)^{-1}dr^2 + r^2d\Omega^2,
 \label{metric-1}
\end{equation}
in which $f(r)$ is an arbitrary $r$-dependent function, ought to
be obtained. The Ricci tensor components, due to the spacetime,
defined by metric (\ref{metric-1}) will be:
\begin{equation}
\begin{array}{l}
R_{00} = \left( -{r}^{2}+2\,Gmr+\frac{1}{3}\,f{r}^{2} \right)\times \\
 \left(
\frac{1}{12}\,{\frac {
 \left( {\it f'}\,{r}^{2}+2\,fr \right) ^{2}}{f}}+
\frac{1}{3}\,\sqrt {3}\sqrt {f{r}^{2}} \left(
-\frac{1}{12}\,{\frac {\sqrt {3} \left( {\it f'}\,{r}^{2}+2 \,fr
\right) ^{2}}{ \left( f{r}^{2} \right) ^{\frac{3}{2}}}}
+\frac{1}{6}\,{\frac { \sqrt {3} \left( {\it f''}\,{r}^{2}
+4\,{\it f'}\,r+2\,f \right) }{ \sqrt {f{r}^{2}}}} \right) {r}^{2}
-\frac{1}{3}\, \left( {\it f'}\,{r}^{2}+2\,f r \right)
r+\frac{1}{3}\,f{r}^{2}
\right) {r}^{-6},\\\\
R_{11}=- \left( \frac{1}{12}\,{\frac { \left( {\it
f'}\,{r}^{2}+2\,fr \right) ^{2}}{f }}+\frac{1}{3}\,\sqrt {3}\sqrt
{f{r}^{2}} \left( -\frac{1}{12}\,{\frac {\sqrt {3}
 \left( {\it f'}\,{r}^{2}+2\,fr \right) ^{2}}{ \left( f{r}^{2}
 \right) ^{\frac{3}{2}}}}
 +\frac{1}{6}\,{\frac {\sqrt {3} \left( {\it f''}\,{r}^{2}+4\,
{\it f'}\,r+2\,f \right) }{\sqrt {f{r}^{2}}}} \right) {r}^{2}
 -\frac{1}{3}\,\left( {\it f'}\,{r}^{2}+2\,fr \right) r+\frac{1}{3}\,f{r}^{2}
\right) {r}^{- 2}\\\times
 \left( -{r}^{2}+2\,Gmr+\frac{1}{3}\,f{r}^{2} \right) ^{-1},\\\\
R_{22}=-\frac{1}{3}\,\sqrt {3}\sqrt {f{r}^{2}} \left(
-\frac{1}{3}\,{\frac {\sqrt {3} \left( {\it f'}\,{r}^{2}+2\,fr
\right) r}{\sqrt
{f{r}^{2}}}}+\frac{1}{3}\,\sqrt {3} \sqrt {f{r}^{2}} \right) {r}^{-2},\\\\
R_{33}=R_{22}\sin^2(\theta).
\end{array}
\label{Ricci}
\end{equation}
Also the Ricci scalar will be derived as:
\begin{equation}
R=\frac{1}{3}\,{\frac {{\it f''}\,{r}^{2}+4\,{\it
f'}\,r+2\,f}{{r}^{2}}}. \label{Ricci scalar}
\end{equation}
Employing these values in the components of Weyl equations in
(\ref{Weyl main}), one obtains:
\begin{equation}
\begin{array}{l}
W_{00}={\frac {1}{324}}\,{r}^{-5}\Big( 72\,{\it f'}\,{r}^{2}-72\,
\left( f \right) r-24\,Gm{{\it f'}}^{2}{r}^ {2}+288\,Gm{\it
f''}\,{r}^{2}-6\,Gm{{\it f''}}^{2}{r}^{4}-432\,r{G}^{2
}{m}^{2}{\it f''}\\-360\,Gm{\it
f'}\,r+360\,{r}^{2}{G}^{2}{m}^{2}{\it f'''} -12\,{{\it
f'}}^{2}{r}^{3}+4\,{f}^{3}r+36\,{\it f''}\,{r}^{3}-36\,{r}^
{5}{\it f''''}-108\,{r}^{4}{\it f'''}\\-3\,{{\it
f''}}^{2}{r}^{5}-120\,G m{\it f'}\,r \left( f \right)
-12\,{r}^{4}Gm{\it f'}\,{\it f'''}-132\, {r}^{3} \left( f
\right) Gm{\it f'''}+96\,Gm \left( f \right) {\it f'' }\,{r}^{2}\\
-48\,{r}^{4}Gm \left( f \right) {\it f''''}-24\,Gm{\it
f'}\,{r}^{3}{ \it f''}-2\,{r}^{5} \left( f \right) {\it
f'''}\,{\it f'}+396\,{r}^{3} {\it
f'''}\,Gm-144\,{r}^{3}{G}^{2}{m}^{2}{\it f''''}\\
+144\,{r}^{4}{\it f''''}\,Gm-4\, \left( f \right) {r}^{4}{\it
f'}\,{\it f''}+12\,{\it f' }\,{r}^{4}{\it f''} +24\,{r}^{5}{\it
f''''}\, \left( f \right) +72\,{r}^{4} \left( f
 \right) {\it f'''}-4\,{r}^{5}{f}^{2}{\it f''''}\\-12\,{r}^{4}{f}^{2}{
\it f'''}+24\,Gm{f}^{2}+4\, \left( f \right) {r}^{3}{{\it
f'}}^{2}-8\, {f}^{2}{r}^{2}{\it f'}+48\,{\it f'}\,{r}^{2} \left( f
\right) -144\,Gm
 \left( f \right) +6\,{\it f'}\,{r}^{5}{\it f'''}\\
 -24\,{r}^{3} \left( f
 \right) {\it f''}+ \left( f \right) {r}^{5}{{\it
 f''}}^{2}-432G^2m^2f'\Big),\\\\
W_{11}=-\frac{1}{36}\,{\frac {1}{{r}^{3} \left( -3\,r+6\,Gm+
\left( f \right) r
 \right) }}\Big(-4\,{r}^{3}{\it f'''}\, \left( f \right) +24\,f+12\,{\it f''}\,{r}^{2}
-24\,{\it f'}\,r\\-4\,{{\it f'}}^{2}{r}^{2}-4\,{f}^{2}+8\,{\it
f'}\,r
 \left( f \right) +2\,{\it f'}\,{r}^{4}{\it f'''}-72\,r{\it f''}\,Gm\\+4
\,{r}^{3}{\it f'}\,{\it f''}-4\, \left( f \right) {\it
f''}\,{r}^{2}- 36\,{r}^{2}{\it f'''}\,Gm-{{\it
f''}}^{2}{r}^{4}+72\,Gm{\it f'}\Big),\\\\
W_{22}=-{\frac {1}{108r^2}}\,\Big(-24\,{\it
f'}\,r-4\,{f}^{2}+8\,{\it f'}\,r \left( f \right) +72\,Gm{ \it
f'}\\-4\, \left( f \right) {\it f''}\,{r}^{2}+2\,{r}^{4} \left( f
 \right) {\it f''''}+4\,{r}^{3}{\it f'''}\, \left( f \right) +4\,{r}^{
3}{\it f'}\,{\it f''}\\+2\,{\it f'}\,{r}^{4}{\it f'''}+12\,{\it
f''}\,{r }^{2}-4\,{{\it f'}}^{2}{r}^{2}-{{\it
f''}}^{2}{r}^{4}-6\,{r}^{4}{\it f''''}\\-12\,{r}^{3}{\it
f'''}-72\,r{\it f''}\,Gm+12\,{r}^{3}{\it f''''}
\,Gm+24\,f\Big)={\frac{W_{33}}{\sin^2(\theta)}}.
\end{array}
\label{Weyl tensor components-1}
\end{equation}
As we expect,
$$W^\alpha_\alpha=0.$$
All the components in (\ref{Weyl tensor components-1}), have a
vacuum solution like:
\begin{equation}
f(r)=-c_1r^2-c_2r-\frac{6GM}{r}. \label{f(r)}
\end{equation}
Substituting (\ref{f(r)}) in (\ref{metric-1}) yields:
\begin{equation}
ds^2 = -\Big(1+\frac{1}{3}{c_2}{r}+\frac{1}{3}c_1r^2\Big)dt^2 +
\Big(1+\frac{1}{3}{c_2}{r}+\frac{1}{3}c_1r^2\Big)^{-1}dr^2 +
r^2d\Omega^2. \label{metric-2}
\end{equation}
Now to evaluate the included constants in (\ref{f(r)}), we shall
use the background field method in the weak field limit. The
zero-zero component of the metric (\ref{metric-1}) can be
rewritten as:
$$g_{00}=\eta_{00}+h_{00},$$
for small fluctuations $h_{00}=\frac{2GM}{r}+\frac{1}{3}f(r)$. The
$r$-component of the Poisson's equation implies that:
\begin{equation}
\nabla^2
h_{00}\equiv(\frac{d^2}{dr^2}+\frac{2}{r}\frac{d}{dr})h_{00}=8\pi
\Big(T_{00}+E_{00}\Big), \label{poisson}
\end{equation}
in which $T_{00}$ is the stress-energy tensor due to the mass of
the source. And here, associated to a charged, spherically
symmetric massive source, the tensor is the volume density:
\begin{equation}
T_{00}=\rho_0=\frac{m_0}{\frac{4}{3}\pi r_0^3}. \label{T00}
\end{equation}
In (\ref{T00}), $m_0$ is the mass of the spherical body and $r_0$
is its known radius. In relation (\ref{poisson}), $E_{00}$ is the
stress-energy tensor, associated to the charge amount of the
massive object. Here, since the source is assumed to be static, we
take the vector potential $A_\mu=(\Phi(r),0,0,0)$, where $\Phi(r)$
is the electric potential at point $r$ in the exterior geometry of
the total charge $q_0$, distributed in a certain volume
($\Phi(r)=\frac{q_0}{r}$). We have \cite{Jackson}:
\begin{equation}
E_{00}=\frac{1}{8\pi}\Big(\frac{q_0}{r^2}\Big)^2+\frac{1}{4\pi}\frac{\partial}{\partial
r}\Big(\Phi(r)\times \frac{q_0}{r^2}\Big) =
\frac{1}{8\pi}\frac{q_0^2}{r^4}. \label{E00}
\end{equation}
Now considering the expression (\ref{f(r)}), and the values in
(\ref{T00}), (\ref{E00}) in (\ref{poisson}), and solving for $c_1$
or $c_2$ yields:
\begin{equation}
c_1 = -3\,{\frac {m_0}{{r_0}^{3}}}-\frac{1}{2}\,{\frac { \left(
q_0
 \right) ^{2}}{{r}^{4}}}-\frac{1}{3}\,\frac{ c_2}{r}.
 \label{c1}
\end{equation}
\begin{equation}
c_2 = -9\,{\frac {rm_0}{{r_0}^{3}}}-\frac{3}{2}\,{\frac { \left(
q_0
 \right) ^{2}}{{r}^{3}}}-3\,{ c_1}\,r.
 \label{c2}
\end{equation}
considering (\ref{c2}) we get:
\begin{equation}
g_{00}^{(1)} =
-(1-\frac{3r^2m_0}{{r_0}^{3}}-\frac{1}{2}\frac{(q_0)^2}{r^2}-\frac{2}{3}c_1r^2).
 \label{g00-1}
\end{equation}
The general spherically symmetric solution to Weyl gravity, has
been derived to be \cite{mannheim2}:
\begin{equation}
g^{W}_{00} =
-(1-\frac{\beta(2-3\beta\gamma)}{r}-3\beta\gamma+\gamma r -kr^2),
\label{general Weyl}
\end{equation}
in which, as it has been mentioned by the authors, the parameters
$\gamma$ and $k$ have been considered to be relevant to the Dark
Energy theory. The term $\frac{2}{3}c_1r^2$ in (\ref{g00-1}),
therefore can be corresponded to the term $kr^2$ in (\ref{general
Weyl}). To estimate a value for $c_1$, we use (\ref{c1}). We
consider the characteristics of the observable universe,
$m_0=m_{\small{obs}}=8\times 10^{55} \,\,\,gr$, and
$r_0=r_{\small{obs}}=4.39\times 10^{28}\,\,\,cm$, which are
respectively, the estimated mass and radius of the observable
universe. We also take $q_0 = q_{\small{obs}} = 0$, because it is
assumed that a finite universe must have a zero net charge
\cite{Landau}. Taking $c_2 =0$ in (\ref{c1}) one obtains:
$$c_1=\frac{3m_{obs}}{{r_{obs}}^3}\approx 2.8\times 10^{-30}\,\,gr/cm^{3},$$
which is comparable to the estimated value for the cosmological
constant (see \cite{Tanhayi} and Ref.s therein).

Now let us consider (\ref{c1}) to obtain:
\begin{equation}
g_{00}^{(2)} =
-(1-\frac{r^2m_0}{{r_0}^{3}}-\frac{1}{6}\frac{(q_0)^2}{r^2}+\frac{2}{9}c_2r).
 \label{g00-2}
\end{equation}
Looking at (\ref{g00-2}), leads us to correspond the term
$\frac{2}{9}c_2r$ to $\gamma r$ in (\ref{general Weyl}). Once more
we use the characteristics of the observable universe in
(\ref{c2}). Taking $c_1=0$ and $r=r_{obs}$ we obtain:
$$c_2=\frac{9M}{{r_{obs}}^2}\approx 0.3736\,\,gr/{cm}^2 \equiv 2\times 10^{-28} \,\,cm^{-1}.$$
And this is exactly the value for $\gamma$ which has been
presented in \cite{mannheim2}, related to the Dark Energy theory.
In comparison with the Reissner-Nordstr\"{o}m-de Sitter metric
\begin{equation}
g_{00}^{RN-d}=-(1-\frac{2m_0}{r}+\frac{(q_0)^2}{r^2}-\frac{1}{3}\Lambda
r^2), \label{RNd}
\end{equation}
the metric (\ref{g00-1}) shows important differences. Specially
when we notice its attractive inverse square potential due to the
charged body, instead of the repulsive one in (\ref{RNd}). Both of
these metrics, have the vacuum energy term, related to the
accelerated expansion of the universe. However, the term in
(\ref{RNd}) would be exactly the cosmological constant, and the
one in (\ref{g00-1}) will have the same value in some limits. On
the other hand, the metric in (\ref{g00-2}) appears to contain
another term, which has been not included by common spherically
symmetric solutions to Einstein field equations. This term, also
by imposing some limits, leads to the same values for the Dark
Energy term, in the general spherically symmetric solutions to
Weyl gravity.

We will continue our discussion, investigating the effective
potentials for the geometrical Black Holes, defined by metrics
(\ref{g00-1}) and (\ref{g00-2}), without the Dark Energy terms.

\section{Effective potentials for a massive charged object around a charged Weyl Black Hole}

Considering a test particle, having the characteristics, $m$ for
mass and $q$ for charge, which is moving on a Weyl Black Hole, one
can derive the effective potential, using the Hamilton-Jacobi
equation of wave crests \cite{Wheeler, Olivares}:
\begin{equation}
g^{\mu\nu}(P_\mu+qA_\mu)(P_\nu+qA_\nu)+m^2=0. \label{H-J}
\end{equation}
$P_\mu$ is the momentum 4-vector\footnote{Here we use the notation
in Ref. \cite{Wheeler}, in which in \S 25.3 the 4-momentum has
been defined like Eq. (\ref{p-1}).}
\begin{equation}
P_\mu=g_{\mu\sigma}P^\sigma=g_{\mu\sigma}\frac{dx^\sigma}{d\lambda}.
\label{p-1}
\end{equation}
where $\lambda$ is the geodesics affine parameter. The metric
components $g_{\mu\nu}$ are derived from the exterior geometry of
the source, namely metric (\ref{g00-1}) and (\ref{g00-2}) without
the cosmological terms. Also the vector potential $A_\mu$ for our
static charged source has been previously defined to be:
\begin{equation}
A_\mu = (\Phi(r),0,0,0), \label{A-mu}
\end{equation}
where $\Phi(r)=\frac{q_0}{r}$ is the scalar electrical potential,
outside the Black Hole. One can define the two conserved
quantities as:
\begin{equation}
E=-P_0, \label{Energy}
\end{equation}
which is the test-particle's energy, and
\begin{equation}
L =P_\phi\,\,\,\,\,\,\, (L\geq0). \label{Angular Momentum}
\end{equation}
which is its angular momentum. We choose $\theta=\frac{\pi}{2}$,
for which the particle's motion is confined to the equatorial
rotations. Therefore
$$P^\theta=\frac{d\theta}{d\lambda}=0.$$
Considering (\ref{g00-1}) in (\ref{H-J}) yields:
$$-\frac{(E-\frac{qq_0}{r})^2}{1-\frac{3r^2m_0}{{r_0}^3}-\frac{1}{2}\frac{{q_0}^2}{r^2}} + (1-\frac{3r^2m_0}{{r_0}^3}-\frac{1}{2}\frac{{q_0}^2}{r^2})^{-1}(\frac{dr}{d\lambda})^2+\frac{L^2}{r^2}+m^2=0,$$
or
\begin{equation}
(\frac{dr}{d\lambda})^2=
(E-\frac{qq_0}{r})^2-(1-\frac{3r^2m_0}{{r_0}^3}-\frac{1}{2}\frac{{q_0}^2}{r^2})(m^2+\frac{L^2}{r^2}).\label{dr-1}
\end{equation}
Equation (\ref{dr-1}) can be rewritten as \cite{Olivares}:
$$(\frac{dr}{d\lambda})^2=[E-\big(\frac{qq_0}{r}-\sqrt{(1-\frac{3r^2m_0}{{r_0}^3}-\frac{1}{2}\frac{{q_0}^2}{r^2})(m^2+\frac{L^2}{r^2})}\,\,\big)][E-\big(\frac{qq_0}{r}+\sqrt{(1-\frac{3r^2m_0}{{r_0}^3}-\frac{1}{2}\frac{{q_0}^2}{r^2})(m^2+\frac{L^2}{r^2})}\,\,\big)],$$
from which we define the effective potential as:
\begin{equation}
V_{eff}^{(1)}=\frac{qq_0}{r}+\sqrt{(1-\frac{3r^2m_0}{{r_0}^3}-\frac{1}{2}\frac{{q_0}^2}{r^2})(m^2+\frac{L^2}{r^2})},
\label{V-eff-1}
\end{equation}
which is the effective potential due to metric (\ref{g00-1}). We
took the positive part to be assured that a positive potential is
available. Using the same procedure for (\ref{g00-2}) yields:
\begin{equation}
V_{eff}^{(2)}=\frac{qq_0}{r}+\sqrt{(1-\frac{r^2m_0}{{r_0}^3}-\frac{1}{6}\frac{{q_0}^2}{r^2})(m^2+\frac{L^2}{r^2})}.
\label{V-eff-1}
\end{equation}
\begin{figure}[htp]
\includegraphics[height=9cm, width=9cm]{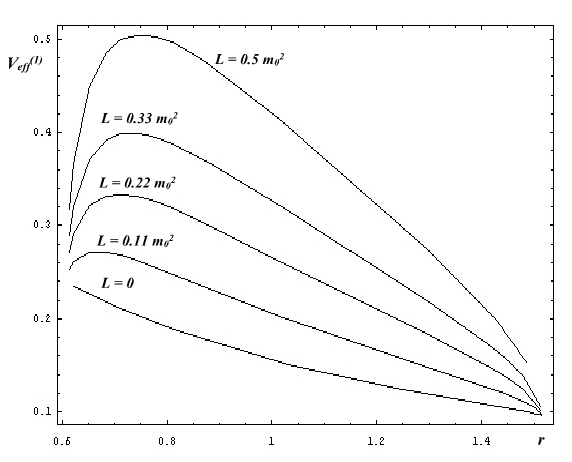}(a)
\hfil
\includegraphics[width=9.2cm,height=9cm]{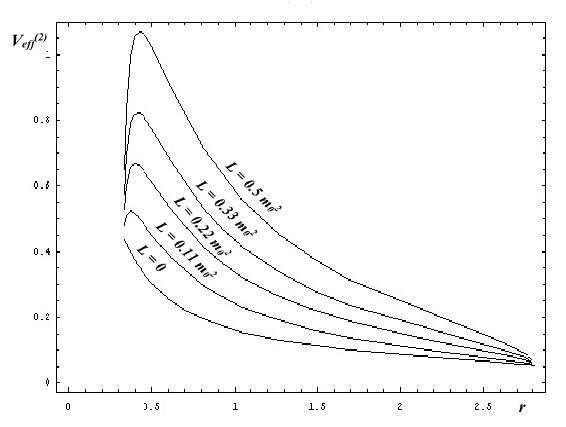}(b)
\hfil
\includegraphics[width=9cm,height=9cm]{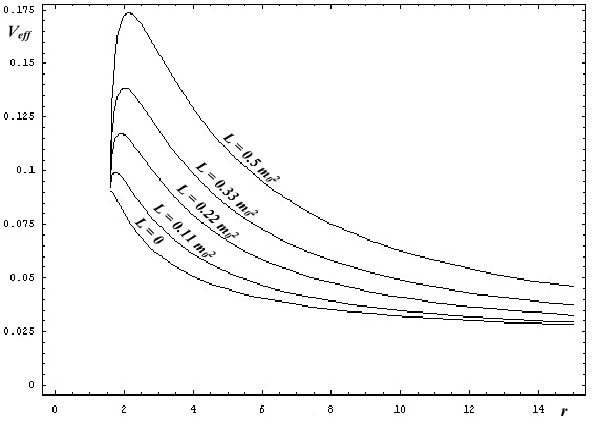}(c)
\caption{\small{(a) The effective potentials for a test-particle,
having different angular momentums, moving on a charged Weyl Black
Hole defined by metric (\ref{g00-1}) and (b) by metric
(\ref{g00-2}). (c) The effective potentials for a test-particle
moving on a Reissner-Nordstr\"{o}m (RN) Black Hole.}}
\label{effective potential-general}
\end{figure}
Figure \ref{effective potential-general} shows illustrations for
these effective potentials for different values of angular
momentums.

As charged Black Holes, either of the Weyl Black Holes must have
two event horizons for $g_{00}^{(1)}=0$ and $g_{00}^{(2)}=0$. For
a Black Hole with a constant radius $r_0$ and mass $m_0$, one
obtains:
\begin{equation}
r_\pm^{(1)}=\frac{1}{6}\,{\frac {\sqrt {6m_{{0}} \left(
{r_{{0}}}^{2}\pm\sqrt {{r_{{0
}}}^{4}-6\,m_{{0}}{q_{{0}}}^{2}r_{{0}}} \right)
r_{{0}}}}{m_{{0}}}}, \label{ev-1}
\end{equation}
\begin{equation}
r_\pm^{(2)}=\frac{1}{6}\,{\frac {\sqrt {6m_{{0}} \left(
3\,{r_{{0}}}^{2}\pm\sqrt {9
\,{r_{{0}}}^{4}-6\,m_{{0}}{q_{{0}}}^{2}r_{{0}}} \right)
r_{{0}}}}{m_{{0 }}}},
 \label{ev-2}
\end{equation}
respectively for (\ref{g00-1}) and (\ref{g00-2}). Note that, for a
Reissner-Nordstr\"{o}m Black Hole we have:
\begin{equation}
r_\pm^{(RN)}=m_{{0}}\pm\sqrt {{m_{{0}}}^{2}-{q_{{0}}}^{2}}.
 \label{ev-RN}
\end{equation}
In the next section, we restrict our discussion to massless
particles.

\section{Effective potentials for massless particles travelling on a Weyl Black Hole}

For massless particles, namely Photon, Neutrino or Graviton, the
characteristics of the test particle changes due to this fact that
the concepts of mass and angular momentum, will break down. We
shall introduce the ratio \cite{Wheeler, Schutz}:
\begin{equation}
b=\lim_{m\longrightarrow0}\frac{L}{(E^2-m^2)^{\frac{1}{2}}}.
\label{b-1}
\end{equation}
Previously we defined:
$$g_{\phi\phi}P^\phi=L\,\,\,\Rightarrow\,\,\,\frac{d\phi}{d\lambda}=\frac{L}{r^2},$$
therefore, according to (\ref{b-1}), from Eq. (\ref{dr-1}) for a
massless neutral particle we have:
\begin{equation}
(\frac{1}{r^2}\frac{dr}{d\phi})^2 -
\frac{1-\frac{3r^2m_0}{{r_0}^3}-\frac{1}{2}\frac{(q_0)^2}{r^2}}{r^2}=
b^{-2}. \label{dr/dphi-1}
\end{equation}
We take
$$C_1^{-2} = \frac{1-\frac{3r^2m_0}{{r_0}^3}-\frac{1}{2}\frac{(q_0)^2}{r^2}}{r^2},$$
for the Weyl Black Hole, which is defined by (\ref{g00-1}), and
also we take
$$C_2^{-2} = \frac{1-\frac{r^2m_0}{{r_0}^3}-\frac{1}{6}\frac{(q_0)^2}{r^2}}{r^2},$$
for the one, defined by (\ref{g00-2}). Note that, for $b\leq C$,
the particle can get to any point $r$. Here, $C^{-2}$ is
considered to be the effective potential for the massless
particles, moving around a Weyl Black Hole. This means that the
maximum, or the critical value for $b$ (we call $b_{crit}$) is the
minimum value for $C$. One can derive this critical value for
either of Weyl Black Holes. For first type Black Holes, this
critical value appears at $r=q_0$. We have:
\begin{equation}
b_{crit}^{(1)}=[C_1]_{min}={\frac {q_0}{\sqrt { \left(
\frac{1}{2}-3\,{\frac {{q_0}^{2}m_0}{{r_0}^{3}}} \right) { }{}}}}.
\label{bcrit-1}
\end{equation}
Also for the second type Weyl Black Holes, the critical point
would be at $r=\frac{q_0}{\sqrt{3}}$, and:
\begin{equation}
b_{crit}^{(2)}=[C_2]_{min}=\frac{\sqrt{3}}{3}\,{\frac {q_0}{\sqrt
{ \left( \frac{1}{2}-\frac{1}{3}\,{\frac {{q_0}^{2}m_0}{{r_0}
^{3}}} \right) {}{}}}}. \label{bcrit-2}
\end{equation}
In Figure 2, the values of effective potentials for a massless
particle for both types of Weyl Black Holes, has been plotted.
\begin{figure}[htp]
\includegraphics[height=9cm, width=9cm]{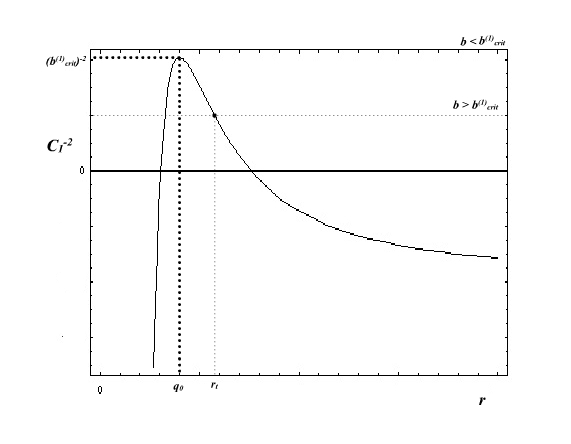}{(a)}
\hfil
\includegraphics[width=9cm,height=9.3cm]{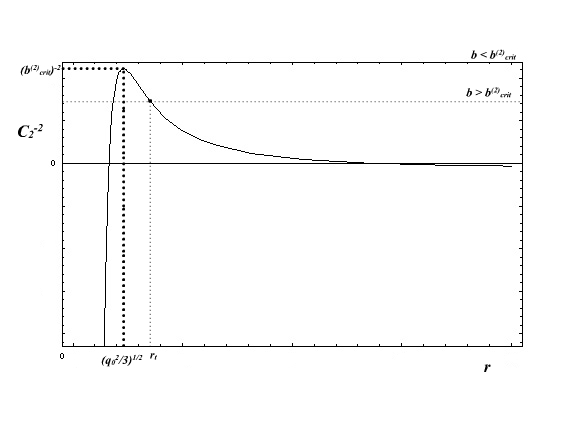}(b)
\caption{\small{(a) The effective potentials for a massless
test-particle, moving on a first type and (b) second type Weyl
Black Hole. $r_t$ would be the turning point. For energies higher
than the $(b_{crit})^{-2}$, the massless particle is captured by
the Black Hole, while for energies lower than this, we have a
reflection from the potential well, having the minimum distance
$r_t.$}} \label{effective potential-massless}
\end{figure}
When $b$ for either of the massless particles on either of Black
Holes, exceeds the $b_{crit}$, then the particle approaches the
Black Hole having the minimum distance $r_t$, and then goes to
infinity. On the maximum of the effective potential, where
$b=b_{crit}$, the particle will have unstable circular orbits. For
$b<b_{crit}$, the test particle which is coming from infinity,
falls into the Black Hole horizon. More details can be found on
text books, for example see \cite{Chandrasakhar}.

\section{Conclusion}

While Einstein theory of relativity, illustrates a finite
classical universe, with a positive acceleration in time like
coordinates, some other gravitational theories, are presenting
solutions for some unexplained features. In this article one of
the most well known ones, namely the Weyl theory of gravity has
been considered. Through this theory, we presented some analytical
expressions for the coefficients, relevant to Dark Energy theory,
and derived their numeric values, which have been in good
agreement with their measured values. Also, we specialized the
spherically symmetric metrics, explaining the exterior geometry of
charged spherical massive source, into two shapes of metric
potentials. These time like metrics, having corresponding
singularities, described two types of charged Black Holes, in
analogy to the Reissner-Nordstr\"{o}m metric. Considering them, we
calculated the effective potentials for massive and massless test
particles and compared them through numerical
illustrations.\\\\\\
\textbf{Acknowledgments}\\
This work was supported under a research  grant by Payame Noor
University.

\end{document}